\documentclass[runningheads]{llncs}

\usepackage[utf8]{inputenc}
\DeclareUnicodeCharacter{03B3}{$\gamma$}

\usepackage{mathptmx}
\usepackage{helvet}
\usepackage{courier}
\usepackage{graphicx}
\usepackage{multicol}
\usepackage[bottom]{footmisc}

\usepackage[dvipsnames]{xcolor}
\usepackage{url}

\usepackage[ruled,vlined,linesnumbered]{algorithm2e}
 \DontPrintSemicolon
 \SetAlFnt{\small}
 
 \SetAlgoCaptionLayout{sppcapsty}
 \newcommand{\sppcommentsty}[1]{\textcolor{black!50}{\sffamily #1}}
 \SetCommentSty{sppcommentsty}
 
 \SetDataSty{sppdatasty}
 
 \SetFuncSty{sppkwfunctiondatasty}

\usepackage{listings}
\usepackage{amsmath}
\usepackage{amsfonts}
\usepackage{amssymb}

\usepackage{hyperref}  
\usepackage[capitalise]{cleveref}

\usepackage[group-separator={\,},binary-units=true,exponent-product = \cdot, output-product = \cdot]{siunitx}

\usepackage{tikz}
\usetikzlibrary{arrows}
\usetikzlibrary{backgrounds}
\usetikzlibrary{shapes.geometric}
\usetikzlibrary{shapes.multipart}
\usetikzlibrary{arrows.meta} 
\usetikzlibrary{decorations.pathmorphing}
\usetikzlibrary{calc}
\usetikzlibrary{shapes}
\usetikzlibrary{patterns}
\usetikzlibrary{decorations.pathreplacing}
\usetikzlibrary{shadows}
\usetikzlibrary{matrix}
\usetikzlibrary{fit}
\usetikzlibrary{scopes}
\usetikzlibrary{hobby}

\graphicspath{{chapter/figures/}}

\usepackage{etoolbox}
\usepackage{ifthen}
\usepackage{xspace}
\usepackage{lipsum}

\usepackage{todonotes}
\usepackage{tabularx}
\usepackage{threeparttable}
\usepackage{booktabs}
\usepackage{subcaption}
\usepackage{fancyvrb}
\usepackage{algorithmic}
\usepackage{multirow}
\usepackage{tikz}

\usepackage{xspace}

\makeatletter

\newcommand{\spp@hc}[1]{\textcolor{Blue}{#1}}
\newcommand{\sppFinalize}{\renewcommand{\spp@hc}[1]{##1}}
\newcommand{\spp@hcs}[1]{\spp@hc{#1}\xspace}
\newcommand{\spp@hcm}[1]{\spp@hcs{\ensuremath{#1}}}

\def\spp@db@funcs{}
\def\spp@db@syms{}
\def\spp@db@abrvs{}
\def\spp@db@protected{}

\newcommand{\spp@push}[2]{%
	\def\spplist{\csname spp@db@#1\endcsname}
	\ifthenelse{\equal{\spplist}{}}{%
		\csgdef{spp@db@#1}{#2}%
	}{%
	  \csgappto{spp@db@#1}{,#2}%
	}
}

\newcommand{\spp@protectcmd}[1]{%
	\spp@push{protected}{#1}%
	\csletcs{spp@copy@#1}{#1}%
}

\newcommand{\NewFunction}[3][Rnd]{%
	\csgdef{#2Sym}{\spp@hcm{#3}}%
	\csgdef{#2Rnd}##1{\spp@hcm{\csname#2Sym\endcsname\left(##1\right)}}%
	\csgdef{#2Sqr}##1{\spp@hcm{\csname#2Sym\endcsname\left[##1\right]}}%
	\csgdef{#2}##1{\csname#2#1\endcsname{##1}}%
	\spp@push{funcs}{#2}%
	\spp@protectcmd{#2}%
	\spp@protectcmd{#2Sym}%
	\spp@protectcmd{#2Rnd}%
	\spp@protectcmd{#2Sqr}%
}

\newcommand{\NewSymbol}[2]{%
	\csgdef{#1}{\spp@hcm{#2}}%
	\spp@push{syms}{#1}%
	\spp@protectcmd{#1}%
}

\newcommand{\NewAbbr}[2]{%
	\csgdef{#1}{\spp@hcs{#2}}%
	\spp@push{abrvs}{#1}%
	\spp@protectcmd{#1}%
}

\newcommand{\sppShowAvailMacros}{%
	\subsubsection*{Function-like}\vspace{-1em}
	\def\tabledata{}
	\foreach \x in \spp@db@funcs {\protected@xappto\tabledata{%
			\expandafter\csname\x\endcsname{x} & \texttt{$\backslash$\x\{x\}} &
			\expandafter\csname\x Sym\endcsname & \texttt{$\backslash$\x Sym}  &
			\expandafter\csname\x Rnd\endcsname{x} & \texttt{$\backslash$\x Rnd\{x\}} &
			\expandafter\csname\x Sqr\endcsname{x} & \texttt{$\backslash$\x Sqr\{x\}} \\
	}}
	\begin{tabular}{|rl|cl|rl|rl|}
		\hline
		\multicolumn{2}{|c|}{Default} & \multicolumn{2}{c|}{Symbol only} & \multicolumn{2}{c|}{Round brackets} & \multicolumn{2}{c|}{Square brackets} \\
		\hline
		\tabledata
		\hline
	\end{tabular}

	\subsubsection*{Symbols}\vspace{-1em}
	\foreach \x in \spp@db@syms {
		\texttt{\textbackslash\x} $\mapsto$ \expandafter\csname\x\endcsname \\
	}

	\subsubsection*{Abbreviations}\vspace{-1em}
	\foreach \x in \spp@db@abrvs {
		\texttt{\textbackslash\x} $\mapsto$ \expandafter\csname\x\endcsname \\
	}
}

\AtEndDocument{
	\foreach \x in \spp@db@protected {
		\ifcsequal{\x}{spp@copy@\x}{}{\PackageError{spp}{macro >\x< redefined}{do not redefine our macros!}}
	}
}

\def\spp@chapter{}
\newcommand{\sppSetupChapter}[1]{%
	\gdef\spp@chapter{#1}%
  \listgadd{\spp@chapters}{#1}%
	\newcounter{spp@authorCount}%
  \setcounter{spp@authorCount}{0}%
	\gdef\spp@chapter{}%
	\gdef\spp@authors{}
}
\newcommand{\sppChapterVar}[1]{\csname spp@chapter@#1\endcsname}

\newcommand{\sppAuthor}[4]{%
	\ifcsdef{spp@insts}{%
		\gdef\newInsts{}%
		\foreach \inst/\mails [count=\i] in \spp@insts {%
			\ifthenelse{\equal{\inst}{#2}}{%
				\xdef\instIdx{\i}%
				\appto{\mails}{,\ \email{#3}}%
			}{}%
			\ifnumgreater{\i}{1}{\xappto{\newInsts}{,}}{}%
			\xappto{\newInsts}{{\expandonce\inst}/{\expandonce\mails}}%
			\xdef\numInsts{\intcalcInc{\i}}%
		}%
	  \ifcsdef{instIdx}{}{%
	  	\xdef\instIdx{\numInsts}%
	  	\gappto{\newInsts}{, {#2}/{\email{#3}}}%
  	}%
	  \let\spp@insts=\newInsts%
	}{%
		\gdef\instIdx{1}%
		\gdef\spp@insts{{#2}/{\email{#3}}}%
	}
	\ifcsempty{spp@authors}{%
		\xdef\spp@authors{\unexpanded{#1\inst}\instIdx}%
	}{%
		\xappto{\spp@authors}{\unexpanded{\and #1\inst}\instIdx}%
	}%
	\ifthenelse{\equal{#4}{}}{}{%
		\appto{\spp@authors}{\orcidID{#4}}%
	}%
	\undef\instIdx%
}

\newcommand{\sppTitle}[2][]{%
	\csgdef{spp@chapter@titlerunning}{#1}%
  \csgdef{spp@chapter@title}{#2}%
  \title{#2}
  \ifthenelse{\equal{#1}{}}{}{\titlerunning{#1}}
}

\newcommand{\sppAbstract}[1]{\gdef\spp@chapter@abstract{#1}}
\newcommand{\sppAuthorRunning}[1]{\authorrunning{#1}}
\newcommand{\sppKeywords}[1]{\gdef\spp@chapter@keywords{#1}}

\newcommand{\sppMakeTitle}{%
	\def\spp@printableInst{}
	\foreach \inst/\mails [count=\i] in \spp@insts {
		\ifnumgreater{\i}{1}{\appto{\spp@printableInst}{\and}}{}
		\xappto{\spp@printableInst}{\expandonce\inst\ \expandonce\mails}
	}
	
	\institute{\spp@printableInst}
	\author{\spp@authors}

	\maketitle%
	\begin{abstract}
		\spp@chapter@abstract
		\keywords{\spp@chapter@keywords}
	\end{abstract}
}

\newcommand{\sppLabel}[2]{\label{#1:\spp@chapter:x-#2}}

\makeatother

\NewFunction{Oh}{\mathcal O}
\NewFunction{OhTil}{\tilde{\mathcal O}}
\NewFunction{oh}{o}
\NewFunction{lTheta}{\Theta}
\NewFunction{lOmega}{\Omega}
\NewFunction{lomega}{\omega}

\NewFunction[Sqr]{prob}{\mathbb P}
\NewFunction[Sqr]{expv}{\mathbb E}
\NewFunction[Sqr]{var}{\texttt{Var}}

\NewSymbol{sN}{\mathbb{N}}
\NewSymbol{sNpos}{\mathbb{N}_{>0}}
\NewSymbol{sZ}{\mathbb{Z}}
\NewSymbol{sR}{\mathbb{R}}
\NewSymbol{NP}{\mathcal{N\mkern-8muP}}

\NewAbbr{etal}{~et~al.}
\NewAbbr{eg}{e.g.,}
\NewAbbr{cf}{cf.} 
\NewAbbr{ie}{i.e.,}

\NewAbbr{Whp}{W.h.p.}
\NewAbbr{whp}{w.h.p.}
\NewAbbr{Wlog}{W.l.o.g.}
\NewAbbr{wLOG}{w.l.o.g.}
\NewAbbr{wrt}{w.r.t.}

\def\SPP{SPP~1736\xspace}

\NewFunction[Sqr]{Text}{\mathit{T}} 
\NewFunction[Sqr]{SA}{\mathit{SA}} 
\NewFunction[Sqr]{SAs}{\mathit{SA}\textnormal{s}} 
\NewFunction[Sqr]{LCP}{\mathit{LCP}} 
\NewFunction[Sqr]{BWT}{\mathit{BWT}}
\NewFunction[Sqr]{BWTs}{\mathit{BWT}\textnormal{s}}
\NewFunction[Sqr]{PFE}{\mathit{PFE}}
\NewFunction[Sqr]{PSEV}{\mathit{PSEV}}
\NewFunction[Sqr]{PSV}{\mathit{PSV}}
\NewFunction[Sqr]{NSV}{\mathit{NSV}}
\NewFunction[Rnd]{Rank}{\operatorname{\mathit{rank}_\alpha}} 
\NewFunction[Rnd]{Select}{\operatorname{\mathit{select}_\alpha}} 

\NewAbbr{ST}{\textit{ST}} 
\NewAbbr{STs}{\textit{ST}s} 
\NewAbbr{PE}{\textnormal{PE}} 
\NewAbbr{PEs}{\textnormal{PEs}} 
\NewAbbr{WT}{\textit{WT}} 
\NewAbbr{WTs}{\textit{WT}s} 
\NewAbbr{WM}{\textit{WM}} 

\sppSetupChapter{nk}
\let\fmtLib\textsc
\newcommand{\nwk}{\fmtLib{NetworKit}\xspace}

\newcommand{\pnwk}{\textsc{jazz}\xspace}
\def\gephi{\fmtLib{Gephi}\xspace}
\def\KADABRA{KADABRA\xspace}

\sppTitle
  {Algorithms for Large-scale Network Analysis \\ and the NetworKit Toolkit} 

\sppAuthor{Eugenio Angriman}{Humboldt-Universit\"at zu Berlin, Germany}{angrimae@hu-berlin.de}{}
\sppAuthor{Alexander van der Grinten}{Humboldt-Universit\"at zu Berlin, Germany}{avdgrinten@hu-berlin.de}{0000-0002-9709-9478}
\sppAuthor{Michael Hamann}{Karlsruhe Institute of Technology, Germany}{michael@content-space.de}{0000-0002-6958-4927}
\sppAuthor{Henning Meyerhenke}{Humboldt-Universit\"at zu Berlin, Germany}{meyerhenke@hu-berlin.de}{0000-0002-7769-726X}
\sppAuthor{Manuel Penschuck}{Goethe University Frankfurt, Germany}{mpenschuck@ae.cs.uni-frankfurt.de}{}

\sppAuthorRunning{E.\,Angriman, A. v. d. Grinten, M.\,Hamann, H.\,Meyerhenke, M.\,Penschuck} 

\sppAbstract{
The abundance of massive network data in a plethora of applications makes scalable analysis
algorithms and software tools necessary to generate know\-ledge from such data in reasonable time.
Addressing scalability as well as other requirements such as good usability and a rich feature set,
the open-source software \nwk has established itself as a popular tool for large-scale network analysis. 
This chapter provides a brief overview of the contributions to \nwk made by the DFG Priority Programme SPP 1736 \textit{Algorithms for Big Data}.
Algorithmic contributions in the areas of centrality computations, community detection, 
and sparsification are in the focus, but we also mention several other aspects -- such as
current software engineering principles of the project and ways to
visualize network data within a \nwk-based workflow.
}

\sppKeywords{
	Network analysis, Algorithms, Software package
}%

\begin{document}%
	\sppMakeTitle
	\section{Introduction}
\sppLabel{sec}{introduction}
\label{sec:nk:introduction}
Network phenomena surround us, be they social contact networks, organizational structures,
or infrastructure networks such as the energy grid, roads or the (physical) internet.
Purely virtual networks such as the world wide web, online social networks, or co-authorship
networks can become particularly large and play an ever increasing role in our daily 
lives~\cite{barabasi2016network,newman2018networks}. Traditional data analysis has been and is
very successful in discovering knowledge from non-network (\eg{} geometric or relational) 
data~\cite{DBLP:books/cu/LeskovecRU14}. Yet, networks and their analysis are about 
``dependence, both between and within variables''~\cite{DBLP:journals/netsci/BrandesRMW13}. To uncover
implicit dependencies hidden in the data, it thus requires appropriate algorithmic techniques
(some of which are also covered in Leskovec\etal's textbook on mining massive datasets~\cite{DBLP:books/cu/LeskovecRU14}).

Massive networks, often with billions of vertices and edges, pose challenges to many established analysis concepts and algorithms due to their prohibitive computational costs.
This leads to the ongoing development of efficient and scalable algorithms. 
The open-source software package \nwk\footnote{\url{https://networkit.github.io/}}~\cite{DBLP:journals/netsci/StaudtSM16} aims to combine a broad range of such algorithms for the analysis of large networks and to make them accessible via consistent, easy to use, and well-documented frontends.
For instance, it offers a feature-rich Python API which integrates into the large Python ecosystem for data analysis.
Under the hood, the heavy lifting is carried out by performance-oriented algorithms that are implemented in C++ and often use multicore parallelism.
The package is also well suited to develop and evaluate novel algorithmic approaches.
As such, \nwk received numerous unique scalable algorithms and implementations in recent years, particularly designed to handle large inputs.

In this chapter, we present a high-level overview of \nwk (\cref{sec:nk:overview})
and portray algorithmic research results derived with and for \nwk\ --
mostly those obtained by projects of \SPP. We cover four main
topics: centrality algorithms (\cref{sec:nk:centrality}), community detection (\cref{sec:nk:community-detection}),
graph sparsification (\cref{sec:nk:sparsification}) as well as graph drawing and network visualization
(\cref{sec:nk:visualization}). While these have been focus areas of \nwk development as part of \SPP, the package has been used in various other application contexts such as 
quantum chemistry~\cite{DBLP:conf/wea/LoozWJM16} and digital humanities~\cite{kreutel2020augmenting}.

\section{\nwk{} --- an Overview}
\sppLabel{sec}{overview}
\label{sec:nk:overview}
\nwk is in development since 2013. The architecture of the current codebase
was released in 2014. At the time of writing, \nwk has a regular release cycle
with two new major releases per year.
Staudt\etal~\cite{DBLP:journals/netsci/StaudtSM16} describe the package's state at the end of 2015.
In this section, we consequently focus on the many additions of new functionality as well as improvements
to the code quality that have been realized in the meantime.
This concerns new performance-oriented graph algorithms,
engineering to speed up existing algorithms, more software engineering guidelines and best practices,
as well as the modernization and extension of \nwk's integration with other tools
within a rich ecosystem (as detailed in \cref{subsec:nk:ecosystem}).

\subsection{Design Considerations}
\label{subsec:nk:design}
\nwk consists of several Python modules wrapping an independently usable core library that is written in C++. Both parts are connected using Cython and are tightly integrated to offer consistent interfaces for most features.
The package is organized into multiple modules, each focusing on one (class of) network analytic problem(s).
Important modules deal with network centrality (\texttt{centrality}), community detection (\texttt{community} and \texttt{scd}) as well as graph generation and perturbation (\texttt{generators} and \texttt{randomization}).
Some novel algorithms in the \texttt{centrality}, \texttt{community}, and \texttt{sparsification} modules that were developed within \SPP are described in more detail in \cref{sec:nk:centrality,sec:nk:community-detection,sec:nk:sparsification}.
Other important modules that are not covered here include
modules for graph algorithms in the language of linear algebra (\texttt{algebraic}, following
the philosophy of GraphBLAS~\cite{DBLP:conf/hpec/KepnerABBFGHKLM16}),
decomposition of graphs into components (\texttt{components}),
distance computations (\texttt{distance}), reading and writing graphs (\texttt{io}),
link prediction (\texttt{linkprediction}), graph coarsening (\texttt{coarsening}), and more.

As a graph data structure, \nwk uses an adjacency array using dynamic arrays (\texttt{std::vector})
to store vertices and their neighborhoods. It also supports edge weights and edge IDs.
This data structure was chosen over static ones such as CSR matrices since it allows for efficient dynamic updates.
The design is complemented by several non-trivial algorithms that can efficiently update their results if the underlying graph changes (\ie after adding and/or deleting edges).

Many of \nwk's algorithms use \fmtLib{OpenMP} for shared-memory parallelism.
In fact, several algorithms in \nwk exhibit best-in-class parallel performance~\cite{DBLP:conf/europar/GrintenAM19}.
Based on an empirical comparison~\cite{DBLP:journals/snam/KochSVM16} between \nwk and several 
distributed frameworks for data and network analysis, \nwk's speed advantage usually remains true 
in comparison to distributed systems with eight-fold resource consumption.
Ref.~\cite{DBLP:journals/snam/KochSVM16} finds that a shared-memory machine
is sufficient to solve many network analytic problems
on real-world instances and concludes that shared-memory parallelism
should be preferred to distributed graph algorithms
as long as the input graph fits into main memory.

\subsection{Ecosystem}
\label{subsec:nk:ecosystem}
In recent years, \nwk matured into an actively maintained open-source project with more than \num{140000} lines of code and a steadily growing number of users and contributors.
By now, the software package exceeds a critical size that warrants efforts beyond the development of new algorithmic features.

To ease contributions and uphold the code quality, \nwk offers detailed guidelines and implements a thorough review process.
We also make heavy use of unit-tests, static code analysis and automated code-formatting as part of our continuous integration pipeline, which targets the three major operating systems.
As many new tests improve the coding standards, we continuously modernize the codebase.
Still, backwards compatibility is a major concern and manifests itself, for instance, in long-term compiler support and in as few changes breaking the API as possible (preceded by a deprecation period of at least one major version release).

Users benefit from a welcoming community, ever-improving documentation, interactive examples showcasing most features, a regular release schedule, and growing support for package managers (currently brew, Conda, pip, and Spack).
\nwk naturally interacts with external projects such as \gephi (see \cref{sec:nk:visualization}), 
\fmtLib{SimExPal}~\cite{DBLP:journals/algorithms/AngrimanGLMNPT19}, and \fmtLib{networkx} as well as graph repositories
and formats including \fmtLib{konect}, \fmtLib{snap}, and \fmtLib{metis};
recent changes make it now even possible to develop standalone \nwk Python modules.

\label{subsec:nk:generators}
Graph data can not only be imported but also be synthesized.
To this end, \nwk offers versatile graph generators in the modules \texttt{generators} and \texttt{randomi\-zation}.
Among others, they are designed to generate and supplement datasets for applications ranging from rapid prototyping to experimental campaigns.
Here, we only mention the supported network \emph{models} since
a different paper surveys graph generation \emph{algorithms} obtained during \SPP.
We include here citations to models or generators developed for/with \nwk.
\begin{itemize}
	\item
		Focus on community structure: Clustered-Random-Graph, LFR, PubWeb, R-MAT, Stochastic Block Model, Watts-Strogatz
	\item
		Prescribed degrees: Havel-Hakimi, Chung-Lu,
		Curveball and Global-Curveball \cite{DBLP:conf/esa/CarstensH0PTW18},
		Edge-Switching 
	\item
		Preferential attachment processes: Barab{\'{a}}si-Albert, Dorogovtsev-Mendes
	\item
		Geometrically embedded: Hyperbolic Random Graph~\cite{DBLP:conf/iwoca/LoozM16,DBLP:journals/jea/LoozM18,DBLP:conf/isaac/LoozMP15,DBLP:conf/hpec/LoozOLM16,DBLP:conf/esa/Blasius0K0PW19},
		Geometric Inhomogenous Random Graph~\cite{DBLP:conf/esa/Blasius0K0PW19},
		Mocnik~\cite{mocnik2018polynomial,DBLP:conf/cosit/MocnikF15}
	\item
		Basic models: $G(n, p)$, Lattice
\end{itemize}
\noindent
Several generators have dynamic variants simulating the evolution of graphs over time.

\section{Centrality Algorithms}
\sppLabel{sec}{centrality-algorithms}
\label{sec:nk:centrality}
One of the most popular concepts used for the analysis of a graph $G = (V, E)$
is \emph{centrality}.
Centrality measures assign a score
to each vertex\footnote{Edge centrality measures are ignored here in the interest of space.} (or group of vertices)
based on its structural position or importance; these scores allow a corresponding
vertex ranking~\cite{DBLP:journals/im/BoldiV14}.
As an example, the well-known Page\-Rank~\cite{DBLP:journals/cn/BrinP12} is a centrality measure originally
devised for web page (and eventually search query) ranking.
It is important to match the underlying research question
with the appropriate centrality measure~\cite{DBLP:series/lnsn/Zweig16} and no single measure is universal.
Thus, dozens of measures have been proposed in the literature~\cite{DBLP:journals/im/BoldiV14}.

As described in more detail below, the centrality research within \nwk revolves not only 
around faster algorithms for computing individual scores and top-$k$ rankings.
Another emphasis is placed on two families of centrality-driven optimization problems
(centrality improvement and group centrality) and how to scale approximation algorithms or heuristics
for their solution to much larger input sizes. For a broader overview, also with a scalability focus,
the reader is referred to Ref.~\cite{ScalingupnetworkcentralitycomputationsAbriefoverview}.

It should also be noted that fast centrality algorithms can be useful in different (but
related) contexts as well; \eg scores of several centrality measures are used
as shortcuts for more expensive influence maximization calculations~\cite{10.1093/comnet/cnz048}.
Also, using score distributions for graph fingerprinting (putting graphs into classes where
all members have similar distributions) is a conceivable use case with the need for numerous
measures that can be computed quickly.

\subsection{Individual Centrality Scores}
\label{subsec:nk:individial-centrality-scores}
We first discuss centrality measures for individual vertices,
\ie measures that assign a centrality score to each $v \in V$.
During \SPP, our focus has been on two classes of centrality measures:
centralities that make use of shortest path computations
(\ie (harmonic) closeness and betweenness)
and algebraic centrality measures
that consider more than just shortest paths
(like Katz centrality and electrical closeness).
\begin{figure}[bt]
\centering
\begin{subfigure}[t]{.33\textwidth}
\centering
\includegraphics[width=\textwidth]{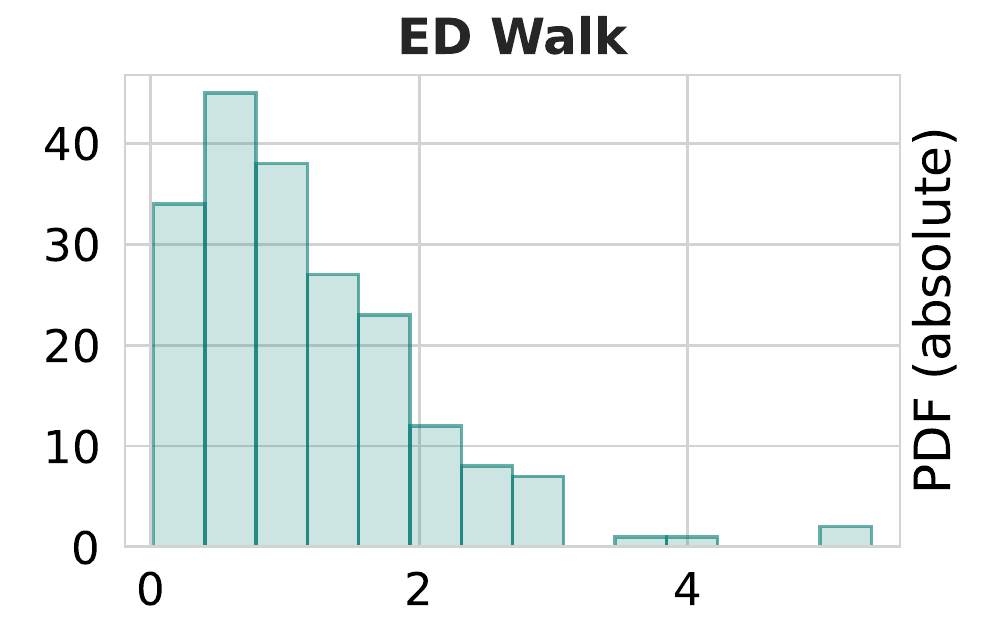}
\end{subfigure}\hfill
\begin{subfigure}[t]{.33\textwidth}
\centering
\includegraphics[width=\textwidth]{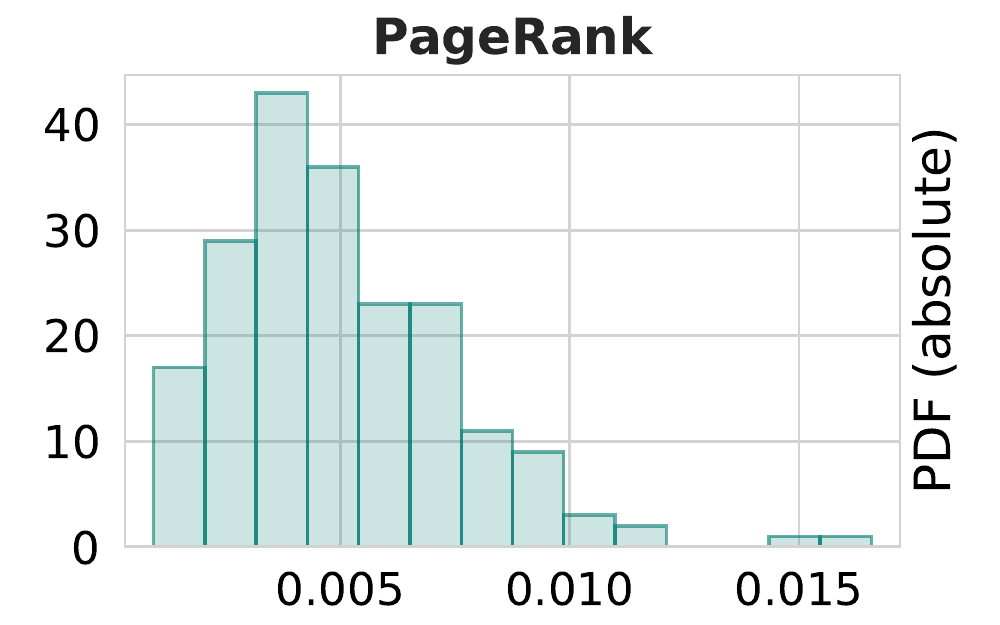}
\end{subfigure}\hfill
\begin{subfigure}[t]{.33\textwidth}
\centering
\includegraphics[width=\textwidth]{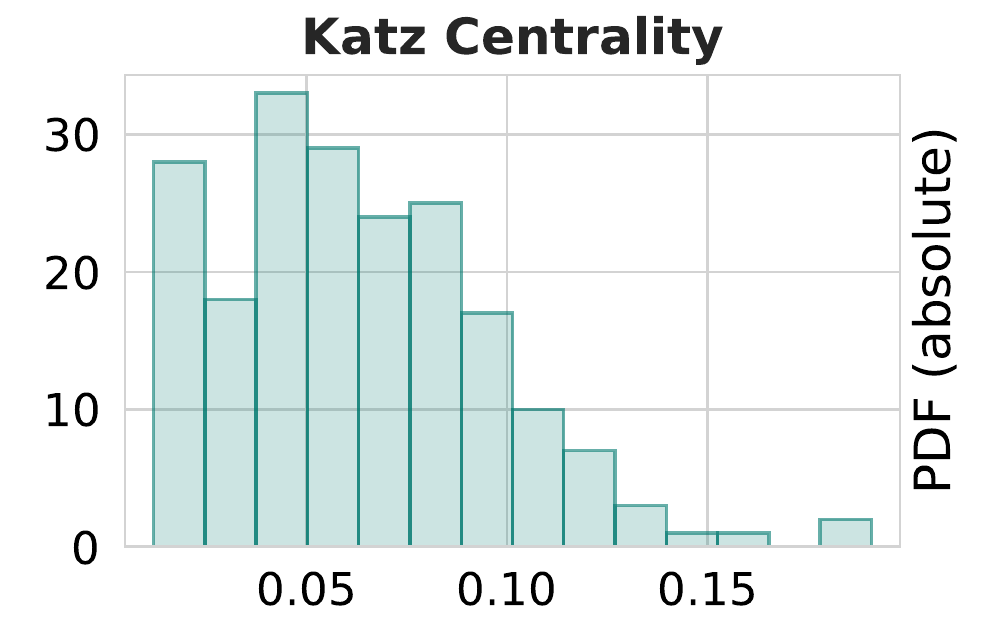}
\end{subfigure}\smallskip

\begin{subfigure}[t]{.33\textwidth}
\centering
\includegraphics[width=\textwidth]{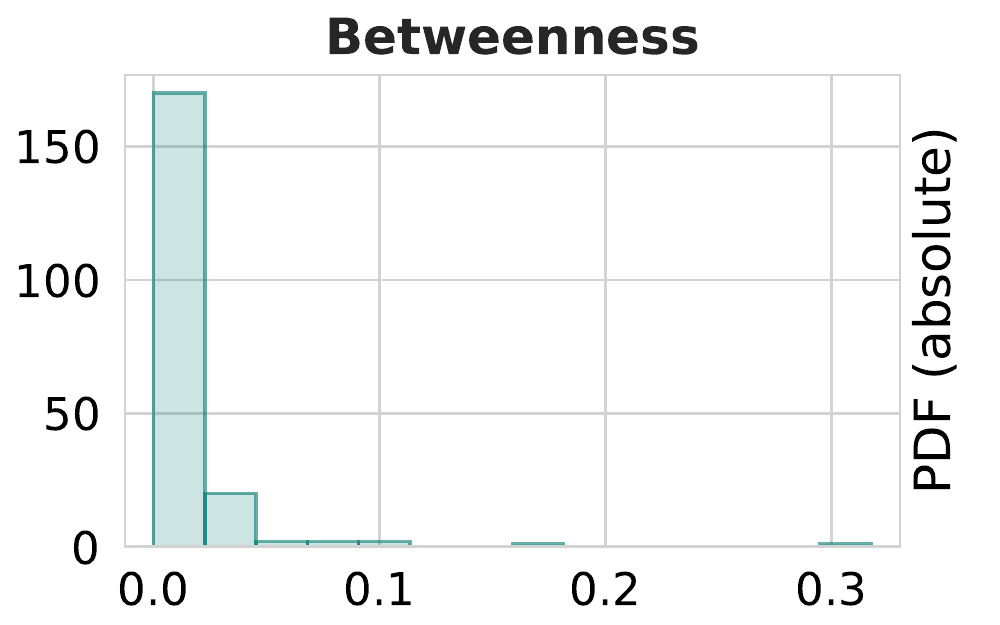}
\end{subfigure}\hfill
\begin{subfigure}[t]{.33\textwidth}
\centering
\includegraphics[width=\textwidth]{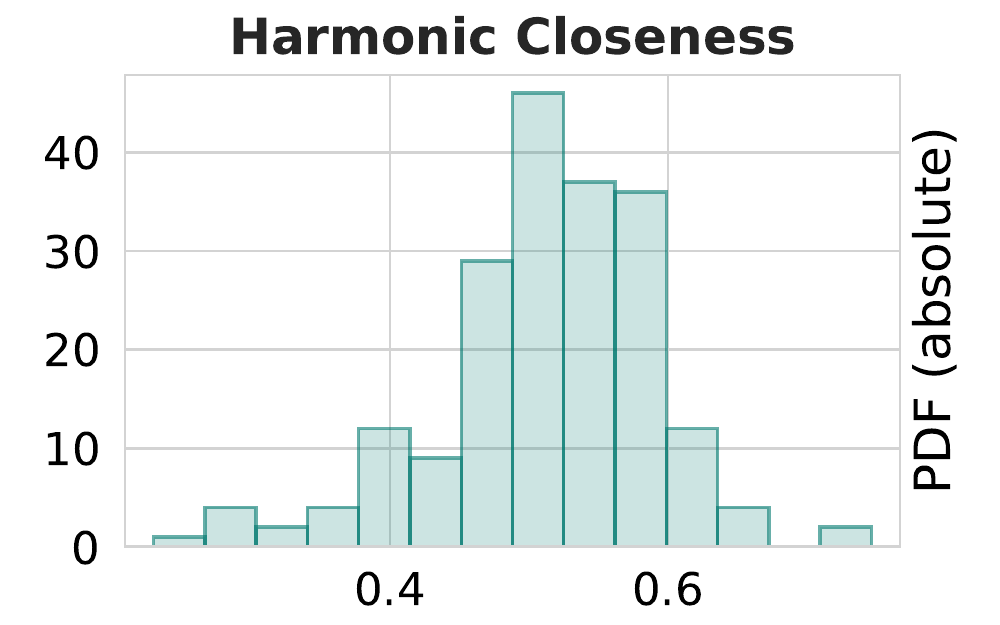}
\end{subfigure}\hfill
\begin{subfigure}[t]{.33\textwidth}
\centering
\includegraphics[width=\textwidth]{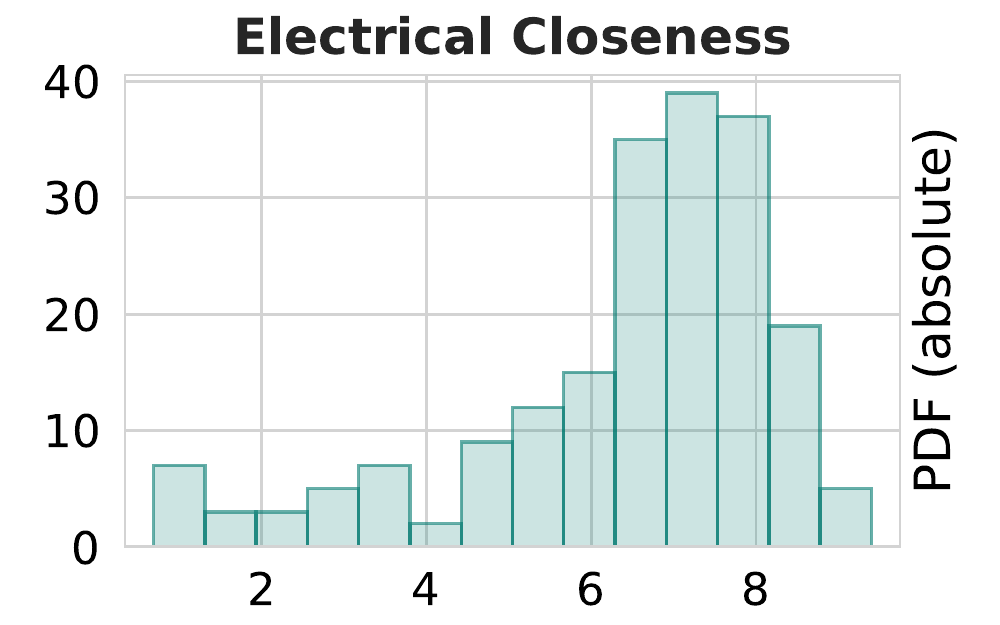}
\end{subfigure}
\caption{Histograms of the distribution of vertex centrality measures of the \pnwk network, which models the collaboration of Jazz musicians~\cite{DBLP:journals/advcs/GleiserD03}.}
\label{fig:nk:prof-centr}
\end{figure}
\cref{fig:nk:prof-centr} depicts the distribution of these centralities
for a single network, including the ED Walk centrality that
we propose in Ref.~\cite{DBLP:conf/alenex/AngrimanGBZGM20}.

\subsubsection{Betweenness.}
\label{subsec:nk:betweeness}

Betweenness centrality is based on the fraction of shortest paths a vertex participates in.
\nwk implements the well-known Brandes algorithm~\cite{doi:10.1080/0022250X.2001.9990249}
for exact betweenness and several algorithms for betweenness approximation.
For static graphs, it has an implementation of the \KADABRA
algorithm~\cite{DBLP:journals/jea/BorassiN19}; additionally, \nwk can approximate betweenness
in dynamic graphs~\cite{DBLP:journals/im/BergaminiM16}. Both of these algorithms employ
a sampling technique that was originally introduced by Riondato
and Kornaropoulos~\cite{DBLP:conf/wsdm/RiondatoK14}.
More precisely, the algorithms sample pairs $(s, t)$ of source
and target vertices uniformly at random. For each
$(s, t)$, a single shortest path is sampled uniformly at random
out of all shortest $s$-$t$ paths. The algorithms
count the number of occurrences of vertices on
these paths; they differ in their stopping conditions.
The multi-threaded
implementation of the static \KADABRA algorithm
additionally exploits a fast data structure
for asynchronous synchronization barriers~\cite{DBLP:conf/europar/GrintenAM19}.
To the best of our knowledge, \nwk's implementation
of \KADABRA is the fastest betweenness approximation code
that is available for multi-threaded machines.

In Ref.~\cite{DBLP:conf/ipps/GrintenM20}, this algorithm was extended to work
with replicated graphs in distributed memory.
The resulting algorithm obtains good parallel speedups
and performs well even on multi-socket shared memory
machines due to the fact that it can avoid NUMA bottlenecks.
Since distributed memory algorithms are outside
the scope of \nwk, this implementation is available externally.


\subsubsection{Closeness.}
\label{subsec:nk:closeness}
Closeness centrality also uses the notion of shortest paths:
it quantifies the importance of a vertex $v\in V$ depending on
how close $v$ is to all the other vertices of the
graph~\cite{bavelas1948mathematical}. It is defined as
$c(v) := (n - 1) / (\sum_{w\in V} d(v, w))$ 
and computing it for a single vertex
requires to run a single-source shortest path (SSSP) algorithm. 
The textbook algorithm to identify the top-$k$
vertices with highest closeness centrality computes $c(v)$ for each vertex of the
graph by running $n$ SSSPs, which is impractical for large-scale networks.
\nwk improves on this by providing an algorithm which finds the top-$k$ vertices with highest closeness
centrality along with their exact value of $c(\cdot)$~\cite{DBLP:journals/tkdd/BergaminiBCMM19}. Even though the
worst-case running time of the algorithm is also $\Omega(|V| |E|)$,
experimental evaluation on real-world data shows that, for small values of $k$,
the algorithm is in practice much more efficient than the textbook
algorithm and other state-of-the-art strategies.

\nwk additionally implements a batch-dynamic version of this
algorithm~\cite{DBLP:conf/alenex/BiseniusBAM18,ABBE2021}, which also addresses harmonic
centrality~\cite{DBLP:journals/im/BoldiV14,rochat2009closeness} -- an alternative definition of closeness
centrality introducing support for disconnected graphs. Experiments on both real-world and synthetic instances
demonstrate that, for moderately large batches of edge updates, the dynamic
algorithm is up to four orders of magnitude faster than a static
recomputation from scratch.


\subsubsection{Electrical Closeness.}
\label{subsec:nk:electrical-closeness}
Electrical resistance is a distance function
on graphs that is constructed by interpreting the
graph as a network of electrical resistors
and by measuring the effective resistance
between vertices in this network.
If the usual distance function (based on shortest-path distances)
in the definition of closeness
is replaced by effective resistance, one obtains the definition
of \emph{electrical} closeness.
This centrality measure has been gaining attention due to the fact
that it considers paths of any length.
\nwk has an efficient approximation algorithm to compute
electrical closeness~\cite{DBLP:conf/esa/AngrimanPGM20}.
This algorithm exploits a
well-known connection
between electrical networks and uniform spanning trees
to approximate electrical closeness faster
than previous numerical algorithms (including
the numerical algorithm from Ref.~\cite{DBLP:conf/siamcsc/BergaminiWLM16})
and can handle graphs with hundreds of millions of edges.

As part of our work on electrical closeness, \nwk
gained support for various numerical algorithms.
These are typically either used as subprocedures of our algorithms
or for performance and/or quality comparisons; however,
they can also be called as standalone numerical solvers.
Experiments with an (in terms of theoretical analysis) fast Laplacian solver 
revealed severe limitations in practice~\cite{DBLP:journals/algorithms/HoskeLMW16}
-- which is why it was discarded.
Instead, we included a fast implementation~\cite{DBLP:conf/siamcsc/BergaminiWLM16}
of the lean algebraic multigrid algorithm (LAMG)~\cite{DBLP:journals/siamsc/LivneB12},
which is particularly well-suited to solve series of Laplacian linear systems with identical system matrices.


\subsubsection{Katz Centrality.}
\label{subsec:nk:katz}
\nwk also implements an approximation algorithm for Katz centrality
that can handle graphs with billions of edges within
a few minutes~\cite{DBLP:conf/esa/GrintenBGBM18}.
The algorithm utilizes lower and upper bounds on the
centrality score of each vertex and improves these bounds
until the Katz centrality ranking is
computed with sufficient precision.
In comparison to earlier combinatorial algorithms
for Katz centrality, our algorithm is the first to
obtain a provable approximation bound
and/or the correctness of the ranking.
It is also at least 50\% faster than numerical methods.
\nwk provides a parallel implementation of this
algorithm that can also handle dynamic graphs.
In Ref.~\cite{DBLP:conf/esa/GrintenBGBM18}, we
additionally provide a GPU-based implementation
which is not part of \nwk.

\subsection{Improving One's Own Centrality}
\label{subsec:nk:improving-own-centrality}
One possible way to improve one's ranking position in a web search is to attract links from influential web pages.
For some time, this led to so-called link farming~\cite{DBLP:books/daglib/0017280} for search engine optimization.
More generally, beyond web search, one wants to increase the centrality of a vertex by adding a specified 
number of new edges incident to it. Crescenzi\etal~\cite{DBLP:journals/tkdd/CrescenziDSV16} addressed this 
problem for closeness centrality.
As a follow-up to that work, Ref.~\cite{DBLP:journals/jea/BergaminiCDMSV18} considered two
betweenness centrality improvement problems: 
maximizing the betweenness \emph{score} of a given vertex (MBI) and
maximizing the \emph{ranking position} of a given vertex (MRI).
The paper proves that both problems are hard to approximate. Unless $\mathcal{P} = \NP$,
MBI cannot be approximated within a factor greater than $1 - \frac{1}{2e}$ and for MRI
there is no $\alpha$-approximation algorithm for any constant $\alpha \leq 1$.
The paper also proposes a simple greedy algorithm for MBI that performs well in practice
and provides a $(1-1/e)$-approximation. This way, MBI can be approximated
for (most) networks with up to $10^5$ edges in a matter of
seconds or a few minutes. The greedy algorithm's implementation builds, among others, upon a dynamic
algorithm for betweenness centrality~\cite{DBLP:conf/wea/BergaminiMOS17} that can update the betweenness 
scores of all vertices much faster after small graph changes (such as the insertion of one or few edges).

\subsection{Group Centrality Optimization}
\label{subsec:nk:group-centrality}
\emph{Group centralities} are network-analytic measures
that quantify the importance of vertex groups~\cite{EB99groupcentrality}.
In contrast to centrality measures that apply to individual vertices,
the goal of these measures is to determine how well
the entire group jointly ``covers'' the graph;
\ie{} the group centrality score is \emph{not} determined
by the scores of individual vertices.

\nwk includes various group centrality algorithms to approximate sets
of vertices that maximize the group centrality score.
Most of the algorithms are based on submodular optimization.
For example, \nwk implements a greedy algorithm to approximate
group degree and the group betweenness maximization
algorithm by Mahmoody\etal~\cite{DBLP:conf/kdd/MahmoodyTU16}.
New algorithms developed as part of \SPP are the
GED-Walk approximation
algorithms from Ref.~\cite{DBLP:conf/alenex/AngrimanGBZGM20} and various group closeness
algorithms; these algorithms are described below.
A very recent addition to \nwk is an approximation algorithm for group forest
closeness centrality; for details we refer to
Ref.~\cite{DBLP:conf/sdm/GrintenAPM21}.

\subsubsection{Group Closeness.}
\label{subsec:nk:group-closeness}
Group closeness measures the importance of
a \emph{group} of vertices $S\subset V$ as the reciprocal of the sum of the
distances from $S$ to the vertices in $V\setminus S$, where the distance from
$S$ to a vertex $v \in V$ is defined by the minimum $d(S, v) := \min_{u \in S}d(u, v)$.
Finding the group $S^\star$ with highest group closeness is known to be an
$\NP$-hard optimization problem~\cite{DBLP:conf/adc/ChenWW16,DBLP:conf/alenex/AngrimanGDGGM21}.
Thus, in practice, the problem is addressed on large-scale networks either with heuristics or
approximation algorithms. \nwk provides a greedy heuristic~\cite{DBLP:conf/alenex/BergaminiGM18} that computes a set of vertices with high
group centrality. On small enough instances where it is feasible to compute the
optimum, it has been shown that the algorithm yields solutions with nearly
optimal quality.

An alternative heuristic, which allows to trade quality for speed, is based on local
search. \nwk implements a family of local search heuristics for group
closeness maximization that achieve different trade-offs between quality and
running time~\cite{DBLP:conf/bigdataconf/AngrimanGM19}. In general, they are one to three orders of
magnitude faster than the greedy algorithm. At the same time, our algorithms retain $80\%$ ---and, in numerous cases, even more than $99\%$--- 
of the greedy algorithm's solution quality. \nwk also includes the first approximation algorithm for
group closeness maximization~\cite{DBLP:conf/alenex/AngrimanGDGGM21} (for
undirected graphs) which yields solutions with higher quality than the greedy
algorithm at the cost of additional running time.

A major limitation of group closeness is that it can only handle (strongly) connected graphs --
the distance between unreachable vertices is either undefined or infinite,
and an infinite denominator results in group closeness score of zero.
Another group centrality measure that also handles disconnected graphs is group harmonic centrality,
which is defined as $GH(S) := \sum_{u\in V\setminus S}d(S, u)^{-1}$.
Maximizing $GH$ has been shown to be an $\NP$-hard
problem~\cite{DBLP:conf/alenex/AngrimanGDGGM21} as well and two approximation
algorithms for group harmonic maximization have been
introduced in Ref.~\cite{DBLP:conf/alenex/AngrimanGDGGM21}; both of them are
available in \nwk.

\subsubsection{GED-Walk.}
\label{subsec:nk:ged-walk}
GED-Walk (GED = group exponentially decaying) is an algebraic group centrality measure
that was introduced in Ref.~\cite{DBLP:conf/alenex/AngrimanGBZGM20}.
Similarly to Katz centrality (which only applies to
individual vertices), GED-Walk counts the number of \emph{walks}
(and not paths) in the graph. Unlike Katz centrality,
it counts walks that \emph{cross} the group of vertices
(instead of counting walks that \emph{start} (or end) at certain vertices).
Computing GED scores can essentially be done via sparse matrix-vector
multiplication; hence, the measure
can be computed faster than centrality measures
that involve the computation of shortest paths.
In Ref.~\cite{DBLP:conf/alenex/AngrimanGBZGM20}, we propose
a greedy algorithm that computes a
group with approximately maximal GED-Walk centrality.
The algorithmic approach is based on techniques derived from our
Katz algorithm~\cite{DBLP:conf/esa/GrintenBGBM18}
and iteratively refines bounds on the group centrality score.
In experiments, GED-Walk maximization turns out to be at least one
order of magnitude faster than the corresponding greedy algorithms for
group betweenness and group closeness.
When applied within semi-supervised vertex classification,
GED-Walk improves the accuracy compared to various existing measures.

\section{Community Detection}
\sppLabel{sec}{community-detection}
\label{sec:nk:community-detection}
Community detection aims to detect subgraphs that are internally densely and externally sparsely connected.
From this fuzzy idea, many formalizations and algorithms have been developed~\cite{OTHER:journals/PhysicsReports/FortunatoH16}.
A division of the graph into disjoint communities is the most frequently studied setting.
The most popular quality measure for this setting is modularity~\cite{OTHER:journal/PhysRevE/NewmanG04}.
As it is $\NP$-hard to find the (clustering with) optimal modularity score~\cite{DBLP:journals/tkde/BrandesDGGHNW08}, heuristics are used in practice. A very popular one is the Louvain algorithm~\cite{DBLP:journals/JoSMTE/BlondelGLL08}.
While it is already quite fast, it is purely sequential in its original formulation
and thus does not exploit the many cores available in modern processors.
Already the earliest work in \nwk{} includes the development of a parallel variant of the Louvain algorithm named PLM~\cite{DBLP:conf/icpp/StaudtM13}.
This first work also includes a fast parallel label propagation algorithm named PLP and an ensemble algorithm that combines several runs of PLP with a final step where PLM is used.
Later improvements to PLM, including the parallelization of additional steps, made PLM so fast that it outperformed the ensemble approach both in terms of speed and quality~\cite{DBLP:journals/tpds/StaudtM16}.
Further, a refinement round similar to Ref.~\cite{DBLP:journals/jea/RottaN11} has been introduced that further increases the quality at the expense of a slightly longer running time.
PLM was later used in a case study on correspondences between clusterings~\cite{DBLP:conf/sdm/GlantzM18}.
With such correspondences one can reveal how one clustering differs from another one, \eg\ when computed
with different algorithms or after minor graph changes.

If only a community around a specific vertex or a set of vertices (so-called seed vertices) is desired, we do not need to detect communities that cover the whole graph.
Many such algorithms greedily add new vertices until a local minimum of a certain quality function is reached.
A first study on such local community detection algorithms~\cite{DBLP:conf/bigdataconf/StaudtMM14} based on \nwk has shown that they are quite slow and imprecise in comparison to PLM.
A more recent study~\cite{DBLP:journals/algorithms/HamannRW17} shows that many local community algorithms detect a community in which the seed is not strongly connected.
Only algorithms that employ further guidance, \eg{} using edge scores based on triangles, are able to correctly identify a community the seed vertex is embedded in.
The study further shows that the results of all local community detection algorithms can be improved by starting with the largest clique in the subgraph induced by the neighbors of the seed vertex.
For this, the possibility to combine two local community detection algorithms has been added to \nwk\ -- a first one that detects the clique and then a second one that expands this clique into a community~\cite{DBLP:journals/algorithms/HamannRW17}.
This allows changing both the seeding strategy and the latter expansion step.

For the experimental evaluation of community detection algorithms, suitable input instances are required~\cite{DBLP:reference/snam/BaderMS0KW14}.
Ideally, instances from applications of community detection with known ground truth communities should be used for this.
However, they are frequently either quite small, unavailable due to privacy concerns or commercial interests, or the available ground truth data cannot be recovered from the graph's structure~\cite{OTHER:journals/PhysicsReports/FortunatoH16,OTHER:journals/corr/PeelLC16}.
For this reason, synthetically generated benchmark graphs with generated ground truth communities are frequently used.
The most popular one is the LFR benchmark graph generator~\cite{OTHER:journals/PhysRevE/LancichinettiFR08}, of which \nwk also provides an implementation for the case of unweighted, undirected graphs with disjoint communities~\cite{DBLP:journals/ans/StaudtHGSM17}. 
Due to a partial parallelization and more efficient data structures, experiments show a speedup compared to the original implementation of 18 to 70 using 16 cores~\cite{DBLP:journals/ans/StaudtHGSM17}.
When the similarity between a detected and a (possible) ground truth community is low, it is often not clear if such a similarity could also be achieved by chance.
Therefore, Hamann\etal~\cite{DBLP:journals/algorithms/HamannRW17} also introduced a simple baseline algorithm using a BFS that stops when the same number of vertices as contained in the ground truth community have been visited and returns them as community.
Together with additional methods for the evaluation of the found communities, \nwk thus provides a comprehensive framework for the development, evaluation, and application of local community detection algorithms.

Nastos and Gao~\cite{DBLP:journals/socnet/NastosG13} suggest quasi-threshold graphs, \ie{} graphs that do not contain a path or cycle of four vertices as vertex-induced subgraph, as a model for communities in social networks.
As a given graph is usually not a quasi-threshold graph, they suggest to insert and delete as few edges as possible to transform a graph into a quasi-threshold graph. The connected components are then considered as communities.
The first scalable heuristic for this problem~\cite{DBLP:conf/esa/BrandesHSW15} has been implemented in \nwk.

\section{Graph Sparsification}
\sppLabel{sec}{graph-sparsification}
\label{sec:nk:sparsification}
Centrality measures suggest that certain vertices or edges are more important than others.
In graph \emph{sparsification}, the idea is to exploit this fact to obtain a subset of the vertices and/or edges that preserve key properties of the graph, \ie{} to select vertices and edges that are important for these properties.
Properties of the graph can be preserved either directly or in a scaled version.
For example, the degree distribution cannot be exactly preserved when we remove edges, but we can preserve the general shape of the degree distribution.
Graph sparsification can provide insights into the structure of a graph, as it provides insights on how much redundancy there is and which edges are important for certain properties.
An application of these insights is speeding up other network analysis tasks or making them possible in the first place by reducing the graph's size such that the running time and memory requirements are reduced~\cite{DBLP:conf/sigmod/SatuluriPR11}.
Further, some of these sparsification techniques can also remove noise from the graph such that, \eg{} more informative drawings can be generated~\cite{DBLP:journals/jgaa/NocajOB15}.
In \nwk, we provide a set of edge sparsification algorithms~\cite{DBLP:journals/snam/HamannLMSW16}.
Given a graph $G = (V, E)$, they identify subsets $E' \subset E$ of the edges such that $G' = (V, E')$ preserves certain properties of $G$.
We currently do not consider vertex sparsification, \ie{} filtering vertices while maintaining properties of the graph
-- since in many network analysis tasks (like vertex centralities or community detection), we are interested in a result for every vertex.
If some vertices were no longer part of the graph, we would need to extrapolate their results, requiring an additional post-processing step for every network analysis task.

\begin{figure}[tb]
  \centering
  \includegraphics[width=0.3\textwidth]{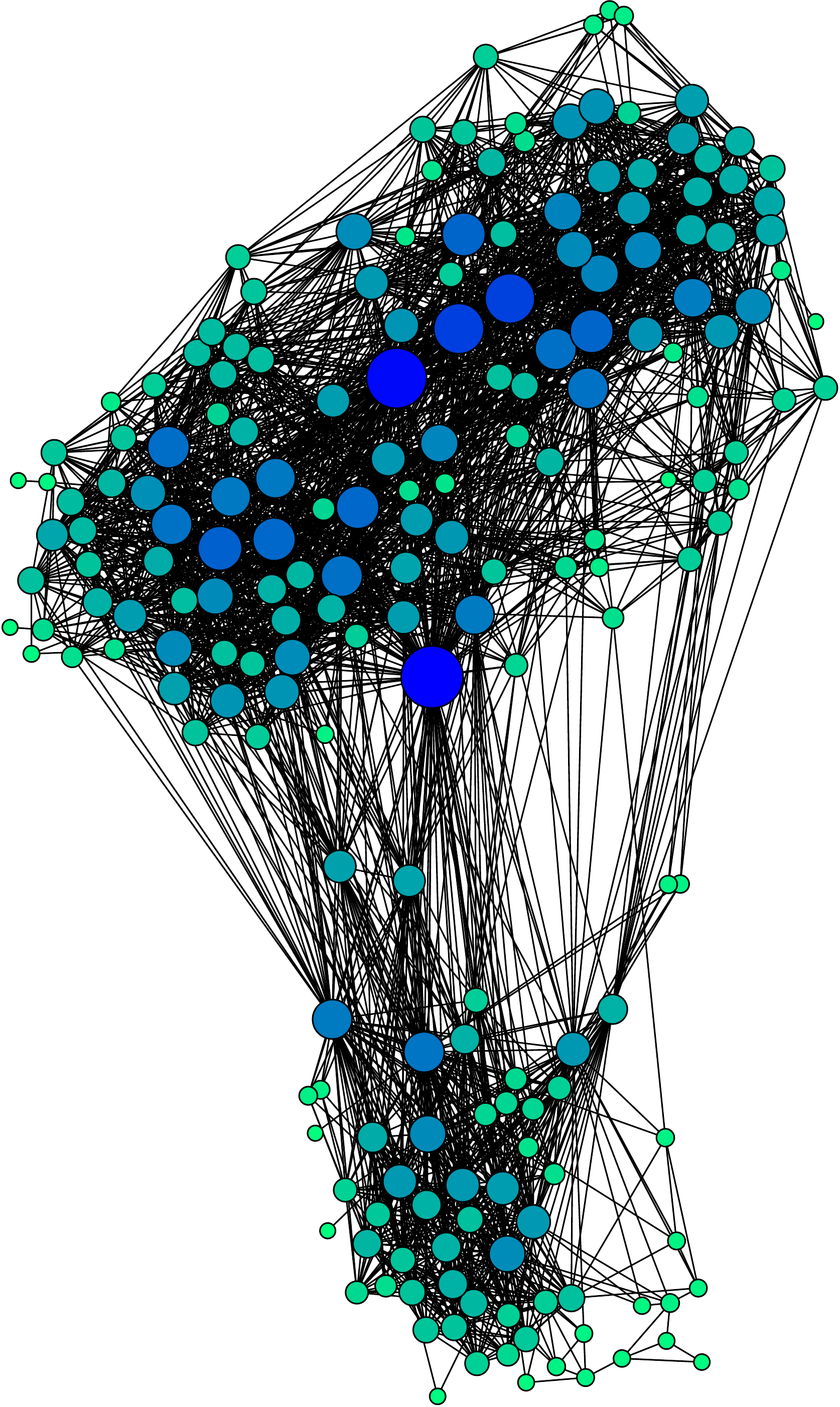}
  \quad
  \includegraphics[width=0.3\textwidth]{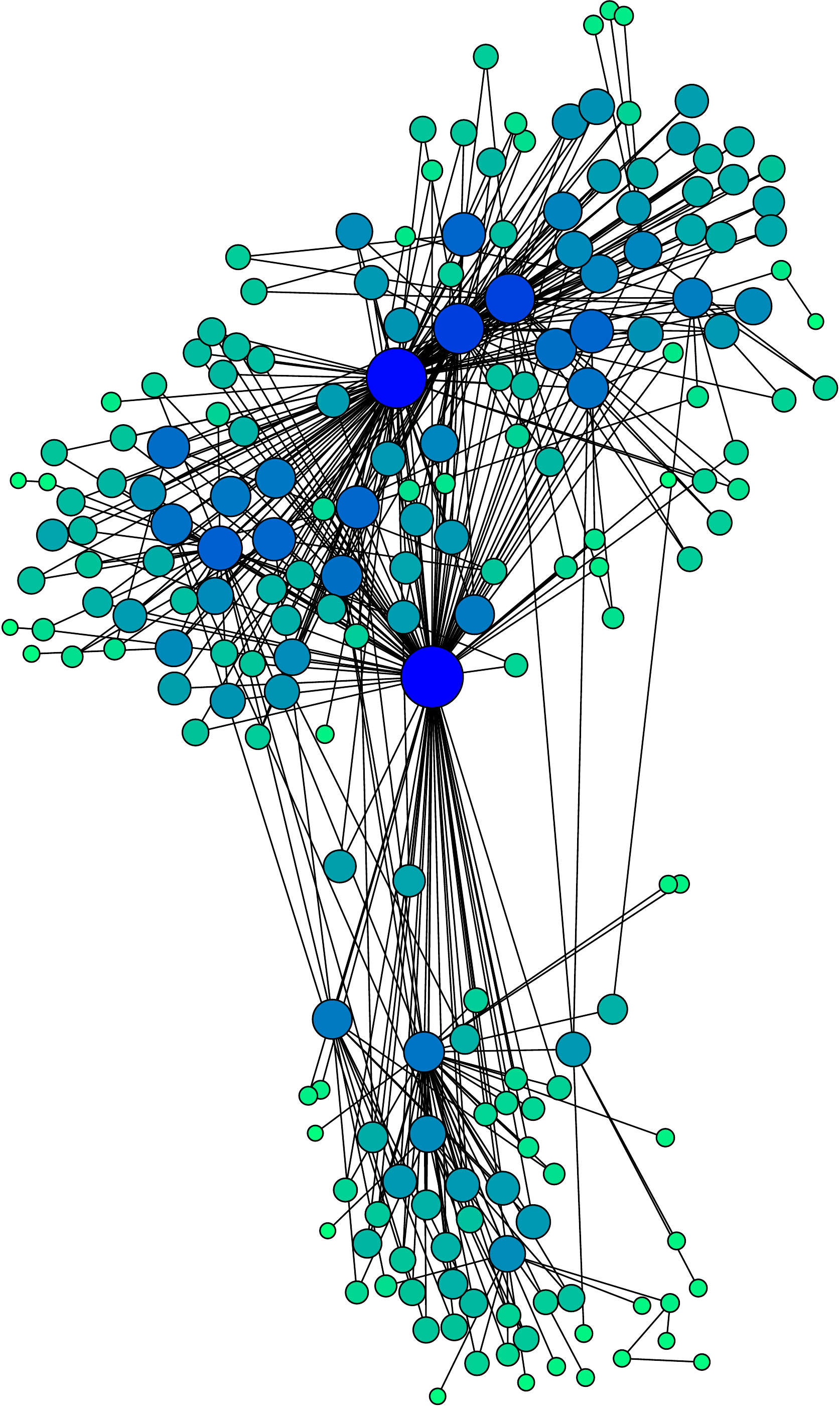}
  \caption{Drawing using \gephi~\cite{DBLP:conf/icwsm/BastianHJ09} of the \pnwk{} network~\cite{DBLP:journals/advcs/GleiserD03} (left) and a sparsified version containing 15\% of the edges (right) using the novel local degree algorithm. Vertex size and color is proportional to degree.}\label{fig:nk:jazz-sparsification}
\end{figure}

With its diverse set of network analysis algorithms, \nwk provides the ideal testbed for sparsification algorithms.
A study~\cite{DBLP:journals/snam/HamannLMSW16} compares a set of six existing and one novel sparsification algorithm as well as five novel variants of the existing algorithms using \nwk.
The study shows that these sparsification algorithms can be classified into three groups: those that primarily preserve edges within densely connected areas, those that primarily preserve connectivity between different areas, and those that are almost or completely random.
The algorithms in the first group strengthen the formation of communities and either keep or increase the average local clustering coefficient as already suggested by previous work~\cite{DBLP:conf/sigmod/SatuluriPR11,DBLP:journals/jgaa/NocajOB15}.
The novel local degree technique, on the other hand, keeps distances in the graph and thus the diameter small, see \cref{fig:nk:jazz-sparsification} for an example.
As the results show, it is also good at preserving vertex centralities.
Completely random filtering also works surprisingly well at preserving a wide range of network properties.
The study shows that all methods perform better for most measures if, instead of directly filtering edges globally, a vertex of degree $d$ keeps its top $d^e$ neighbors for some exponent $e < 1$.
This local filtering step has been proposed before~\cite{DBLP:conf/sigmod/SatuluriPR11} for a single sparsification algorithm and the study suggests to apply it to all considered algorithms.
In particular, this preserves connectivity of the graph quite well and in general leads to a more even distribution of the preserved edges.

All of these sparsification algorithms can be decomposed into two steps: A first step that assigns each edge a score and a second step that only keeps a certain fraction of the highest-rated edges.
Even the local filtering step can be implemented as a transformation of edge scores.
This makes it possible to easily combine existing and new algorithms.
Further, the resulting scores can be considered as edge centrality measures that permit a ranking of the edges.
With the help of visualization software like \gephi~\cite{DBLP:conf/icwsm/BastianHJ09} (\cref{sec:nk:visualization}), the scores can also be visualized or used for interactive filtering of edges.

\section{Graph Drawing and Network Data Visualization}
\begin{figure}[tb]
\centering
\includegraphics[width=.95\textwidth]{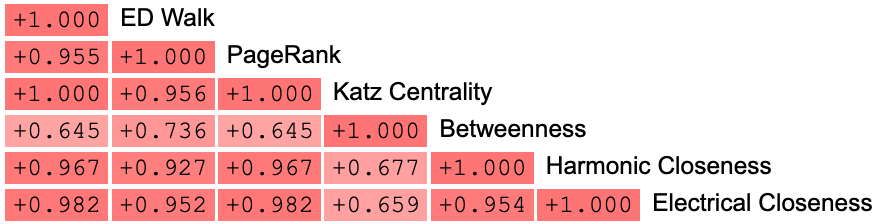}
\caption{Spearman's correlation coefficients between vertex rankings obtained with
different centrality measures for the \pnwk network. Darker [lighter] block shades
indicate higher [smaller] correlation values.}
\label{fig:nk:prof-corr}
\end{figure}

\sppLabel{sec}{graph-drawing}
\label{sec:nk:visualization}
In exploratory network analysis, one needs to evaluate several properties of the network,
which requires writing code to run algorithms and plot their results.
To speed up this process, \nwk provides a dedicated \texttt{profiling} module that allows non-expert users
to run several network analysis algorithms as a single program and visualize
their results in a graphical report that can be rendered in a Jupyter Notebook
or exported as an HTML or a \LaTeX\ document.
As thoroughly explained in Ref.~\cite{DBLP:journals/netsci/StaudtSM16}, first the report
lists global properties of the networks such as the size and the density. Then
it provides an overview of the distribution of several centrality networks as
histograms (as shown in \cref{fig:nk:prof-centr}, \cref{sec:nk:centrality}),
followed by a more detailed statistical analysis. Finally, the report
includes a matrix with the Spearman correlation coefficients between the
rankings of the vertices according to the considered centrality measures; an
example for the \pnwk network is shown in Figure~\ref{fig:nk:prof-corr}.

\begin{figure}[bt]
\centering
\includegraphics[width=.6\textwidth]{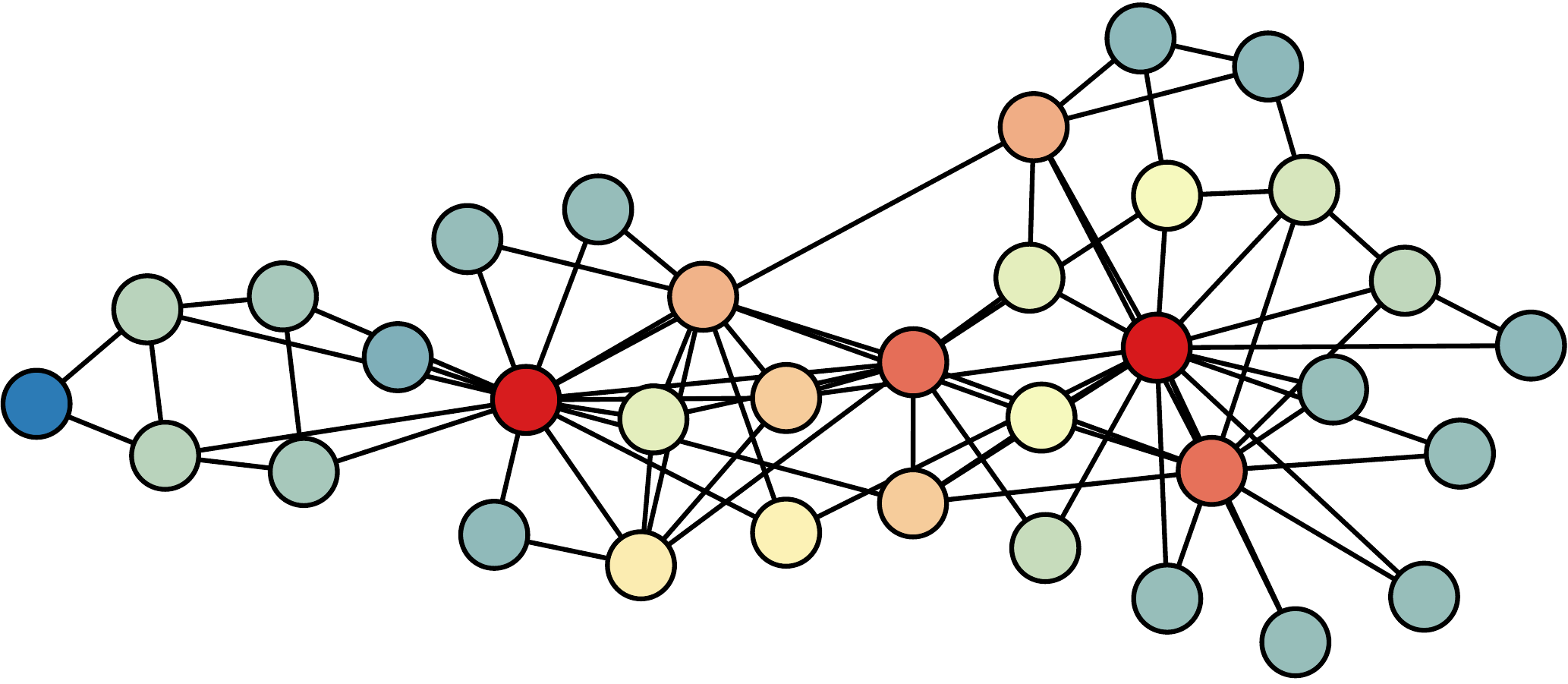}
\caption{Visualization example with \gephi of the \textsc{karate} graph.
Red vertices have the highest harmonic centrality, blue vertices the lowest.}
\label{fig:nk:gephi}
\end{figure}

When dealing with large graphs, statistical overviews as the ones mentioned are indispensable,
since the well-known vertex-edge diagrams do not even scale to graphs of medium size (without further
adjustments). For small graphs, however, visualizations such as those diagrams can be very
valuable. In general, the goal of graph visualization~\cite{DBLP:books/ph/BattistaETT99} is to represent
graphs in a form that is meaningful to the human eye.
Popular application areas for graph visualization are biology (\eg genetic maps),
chemistry (\eg protein functions)~\cite{DBLP:journals/tvcg/HermanMM00}, social network analysis~\cite{kreutel2020augmenting},
and many more.
\gephi~\cite{DBLP:conf/icwsm/BastianHJ09} is a popular Java-based GUI application to explore and visualize graphs.
\nwk's \texttt{gephi} module~\cite{DBLP:journals/snam/HamannLMSW16} allows to
use \gephi to visualize graphs along with additional vertex- or edge
attributes with minimal effort.
Figure~\ref{fig:nk:gephi} shows the visualization in \gephi of the
popular \textsc{karate} graph obtained by the ForceAtlas2 graph drawing
algorithm~\cite{jacomy2014forceatlas2} and by coloring the vertices according
to their harmonic centrality score.

Graph drawing actually precedes visualization in most cases. It is the process of computing
meaningful coordinates for the graph vertices where such information is not supplied with
the graph. \nwk's approach for the most part is to use the graph
drawing capability in \gephi. It has, however, also an implementation of an algorithm
for the maxent-stress objective function, following Ref.~\cite{DBLP:journals/tvcg/MeyerhenkeN018}.
Here, the main intention is to solve an optimization problem that computes
the three-dimensional structure of biomolecules, given distance information between some atom pairs.
To this end, the original algorithm received several application-specific adaptations~\cite{DBLP:conf/esa/WegnerTSM17},
\eg to be able to handle noisy data appropriately.
As a result, the new algorithm by far outperforms its competitors in terms of speed and flexibility, and  often even produces a superior solution quality.

\section{Conclusions}
\sppLabel{sec}{conclusions}
\label{sec:nk:conclusions}
The main design goals of \nwk (speed, rich feature set, usability, and integration into an ecosystem) 
prove to be very useful for users, but they can also be challenging for the developers. One lesson learned
to keep an academic open-source project of this size manageable and alive, is to combine best practices in both 
software engineering and algorithm engineering~\cite{DBLP:journals/algorithms/AngrimanGLMNPT19}.
For example, a proper modularization allows easier reuse and combination of components, leading to a
better extensibility and maintainability. These keywords are well-known in software engineering,
but they also have their effect in algorithm design and implementation -- in particular a simplified
exploration of the design space in experimental algorithmics. \nwk has already proved to be very useful
in this respect for developers.

We have seen that approximation and parallelism can bring us a long way regarding scalability.
They are the obvious, but certainly not the only choices for acceleration: exploiting the 
structure of the data, \eg\ small vs.\ large diameter~\cite{DBLP:journals/tkdd/BergaminiBCMM19}, 
can yield significant speedups on real-world data --- even in the context of exact computations
and potentially on top of parallelism.

\nwk is constantly improved and extended -- according to the resources available to the project.
There are numerous ideas for larger updates from various angles -- of which we mention only two
representative ones: inherent support for attributes within (some of) the algorithms and 
further/improved integration with other tools. The latter is particularly geared towards a closer
connection with machine learning, both on an algorithmic and a software tool level. Given the current
interest in machine learning for data analysis, complete workflows within one seamless toolchain
including \nwk and tools such as \fmtLib{scikit-learn} can be expected to be very attractive for
users from many domains.

	\bibliographystyle{splncs04-spp}
	\bibliography{chapter/references.bib}
\end{document}